\documentclass[aps, prl, twocolumn, a4paper, floatfix, 11point]{revtex4-1}

\usepackage[utf8]{inputenc}
\usepackage{epsfig}
\usepackage[T1]{fontenc}
\usepackage{t1enc}
\usepackage{graphicx}
\usepackage{amssymb}
\usepackage{amsmath}

\usepackage{bm}

\usepackage{color}

\newcommand{\avg}[1]{{\left<#1\right>}}

\usepackage{mathtools}
\def\multiset#1#2{\ensuremath{\left(\kern-.3em\left(\genfrac{}{}{0pt}{}{#1}{#2}\right)\kern-.3em\right)}}

\begin{document}

\title{No need for conspiracy: Self-organized cartel formation in a
modified trust game}

\author{Tiago P. Peixoto}
\email{tiago@itp.uni-bremen.de}
\affiliation{Institut f\"{u}r Theoretische Physik, Universit\"at Bremen,
Otto-Hahn-Allee 1, D-28359 Bremen, Germany}
\author{Stefan Bornholdt}
\email{bornholdt@itp.uni-bremen.de}
\affiliation{Institut f\"{u}r Theoretische Physik, Universit\"at Bremen,
Otto-Hahn-Allee 1, D-28359 Bremen, Germany}


\begin{abstract}
We investigate the dynamics of a trust game on a mixed population where
individuals with the role of buyers are forced to play against a
predetermined number of sellers, whom they choose dynamically. Agents
with the role of sellers are also allowed to adapt the level of value
for money of their products, based on payoff.  The dynamics undergoes a
transition at a specific value of the strategy update rate, above which
an emergent cartel organization is observed, where sellers have similar
values of below optimal value for money. This cartel organization is not
due to an explicit collusion among agents; instead it arises
spontaneously from the maximization of the individual payoffs.  This
dynamics is marked by large fluctuations and a high degree of
unpredictability for most of the parameter space, and serves as a
plausible qualitative explanation for observed elevated levels and
fluctuations of certain commodity prices.
\end{abstract}

\maketitle

Modern societies are complex systems, where observed macroscopic
properties are the emergent result of collective actions of individual
agents. The most commonly adopted scenario assumes that agents select
strategies which are perceived to be in their own best interest. This
decision, however, must often be made without full \emph{a priori}
knowledge of the most likely outcome, and thus must rely on some notion
of belief, or trust. Common examples include most types of markets,
where buyers must decide if a certain product is worth its cost, and
sellers must decide which price they should assign for their
products. At the most fundamental level, this problem can be framed as a
trust game~\cite{berg_trust_1995, mccabe_positive_2003,
sigmund_calculus_2010}, where a buyer must decide whether he buys a
product at a given cost, and the seller decides which cost to select. If
the price is perceived to be fair by both parties, the outcome is
positive for both of them, otherwise it slants in favor of either
party. A real market, however, is composed of many buyers and sellers,
and depending on the situation, buyers do not have the option of not
buying, instead they can only realistically choose \emph{from whom} they
buy. This is often the case, for instance, for car owners who must buy
gasoline, people who must buy groceries, bank account and credit card
owners, etc. In this Letter, we investigate the dynamics of a trust game
on a population of agents who face this restriction, and form an
adaptive network of interactions~\cite{zimmermann_cooperation_2005,
pacheco_coevolution_2006, szabo_evolutionary_2007, li_how_2007,
gross_adaptive_2008, lee_emergent_2011, lee_cooperation_2011,
zhang_coevolving_2011, ben-naim_structure_2006}. We identify the
emergence of an effective cartel-like dynamics, where agents share low
values of value for money, to the overall benefit of the sellers and
detriment of the buyers. This cartel dynamics emerges without any
explicit collusion among the agents, who react independently in order to
maximize their payoff. In this dynamical phase, the evolution of the
average value for money in the population is marked by very large
fluctuations, and high degree of unpredictability, with aperiodic
behavior and very broad spectral densities. These variations are a
result of a never-ending tug-of-war between sellers and buyers, where
buyers seek the best sellers, who in turn compete among themselves,
while at the same time benefiting collectively from uniformly low value
for money. This type of dynamics can be directly compared to the time
evolution of certain commodity prices such as gasoline, which is known
to fluctuate considerably between gas
stations~\cite{eckert_empirical_2011, maskin_theory_1988,
noel_edgeworth_2007}, both in space and time, sometimes with multiple
price changes within a single day, without any apparent connection to
the fluctuation of crude oil prices. Our model provides a conceptual
explanation of the origin of such fluctuations, which does not require
the explicit collusion among the sellers as a necessary element driving
price changes.

Our model is defined as follows. We consider a population of $N$ agents,
where each agent has two simultaneous roles: donator (e.g. a buyer) and
rewarder (e.g. a seller). To each agent $i$ we assign a value for money
variable $w_i \in [0, 1]$. This can be interpreted, for instance, as the
quality of a sold product or service. Each agent is forced to choose
exactly $K$ rewarders to whom it must donate~\footnote{We considered
also the variant where the number of rewarders is randomly distributed
according to a Poisson with average $K$. The results were qualitatively
very similar, with the only noticeable deviation being slightly
different critical values $a_c$ marking the dynamical phase
transition.}. This forms a network of $N$ nodes, where the adjacency
matrix $A_{ij}$ describes the donators' choices. Each agent $i$ has a
donator and a rewarder payoff, $P^+_i$ and $P^-_i$, respectively,
defined as,
\begin{align}
  P^+_i &= \sum_j A_{ij} w_j \label{eq:p+} \\
  P^-_i &= (1-w_i) k_i, \label{eq:p-}
\end{align}
where $k_i = \sum_j A_{ji}$ is the number of donators who choose agent
$i$ (the in-degree of $i$). Eq.~\ref{eq:p+} can be interpreted simply as
the overall satisfaction a customer has with his buying choices, and
Eq.~\ref{eq:p-} as the overall profit a business makes, which is assumed
proportional to how many customers it has, $k_i$, and to the complement
of the value for money it provides, $1-w_i$. We assume these payoff
values correspond to continued interactions between players, instead of
single isolated events (e.g. repeated games after an unspecified number
of rounds), and are accumulated on a time scale which is much faster
than the strategy update dynamics. The strategies of each agent
correspond to their chosen value for money $w_i$, and their choice of
rewarders. These strategies are updated dynamically as follows. At each
time step, a agent $i$ is randomly chosen. With probability $a$ its
rewarder strategy is updated, otherwise its donator strategy is
updated. The actual strategy updates are performed according to the
following rules:
\begin{enumerate}
  \item \emph{Donator update}. For the agent $i$, a random, currently
        chosen rewarder $j$ is selected, so that $A_{ij} = 1$, and
        compared with another rewarder $l \ne j$, with $A_{il} = 0$,
        randomly chosen among the entire population. If $w_l \ge w_j$,
        then the rewarder is replaced, i.e. $A_{ij} \to 0$ and $A_{il}
        \to 1$.
  \item \emph{Rewarder update}. For the agent $i$, another
          agent $j \ne i$ is randomly selected from the population. If
          $P^-_j \ge P^-_i$, then its value for money is copied,
          i.e. $w_i \leftarrow w_j$.
\end{enumerate}
Thus the donator strategies are updated by simple comparison, and the
rewarder strategies are updated by replication. This is so chosen, since
it is always better for a donator to switch to a better rewarder,
whereas it is not \emph{a priori} obvious which is the best value for
money a rewarder should select: If the value $w$ for a given agent is
lowered, the higher will be its payoff immediately, but on the other
hand, the larger is the likelihood it will lose donators as soon as they
update their strategies. Conversely, if the value of $w$ is increased,
it will decrease the rewarders payoff immediately, but it may attract
more donators in the future, which will then cause an increased
payoff. Replication based on payoff automatically chooses the strategies
which are more successful at a given stage, and is thus the most
commonly adopted scenario in evolutionary game
theory~\cite{sigmund_calculus_2010}.

\begin{figure}[hbt!]
  \begin{minipage}{0.49\columnwidth}
    \centering
    \includegraphics[width=\columnwidth]{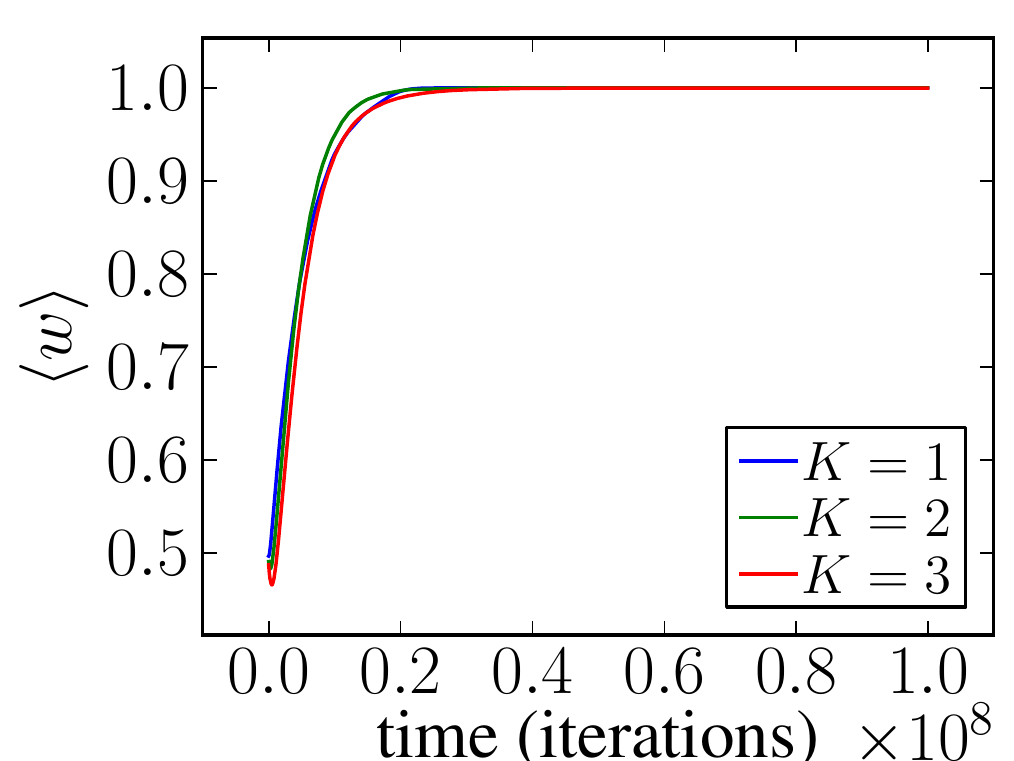}\\
    (a) $a=0.1$
  \end{minipage}
  \begin{minipage}{0.49\columnwidth}
    \centering
    \includegraphics[width=\columnwidth]{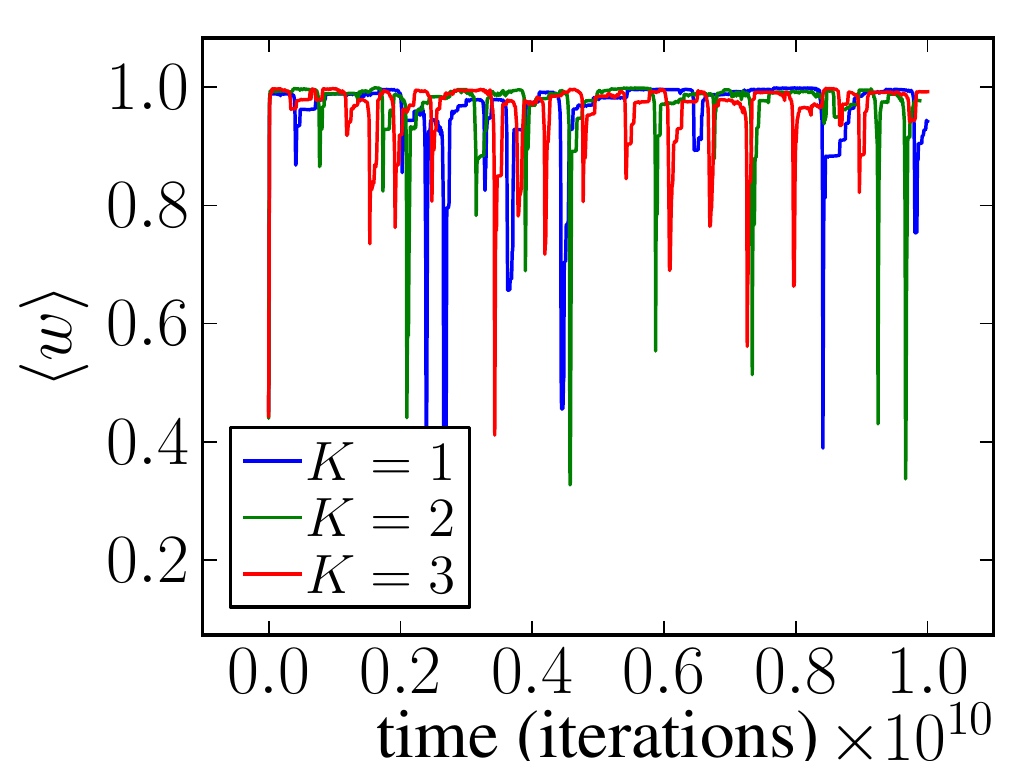}\\
    (b) $a=a_c$
  \end{minipage}
  \begin{minipage}{0.49\columnwidth}
      \includegraphics[width=\columnwidth]{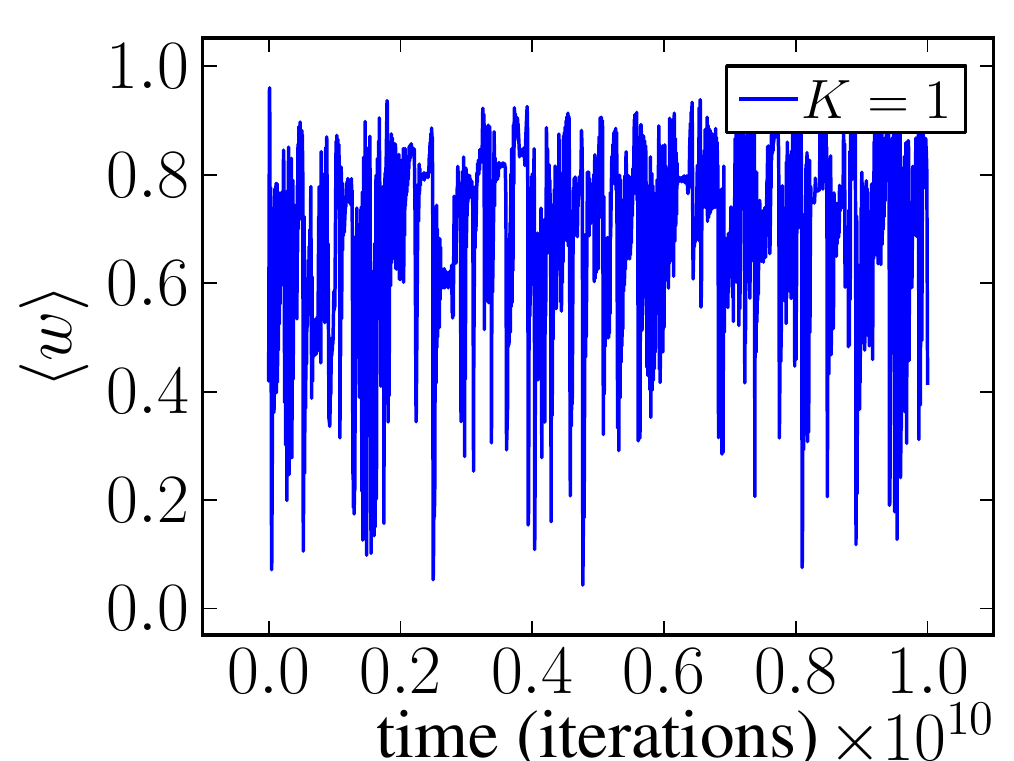}\\
    (c) $K=1, a > a_c$
  \end{minipage}
  \begin{minipage}{0.49\columnwidth}
    \includegraphics[width=\columnwidth]{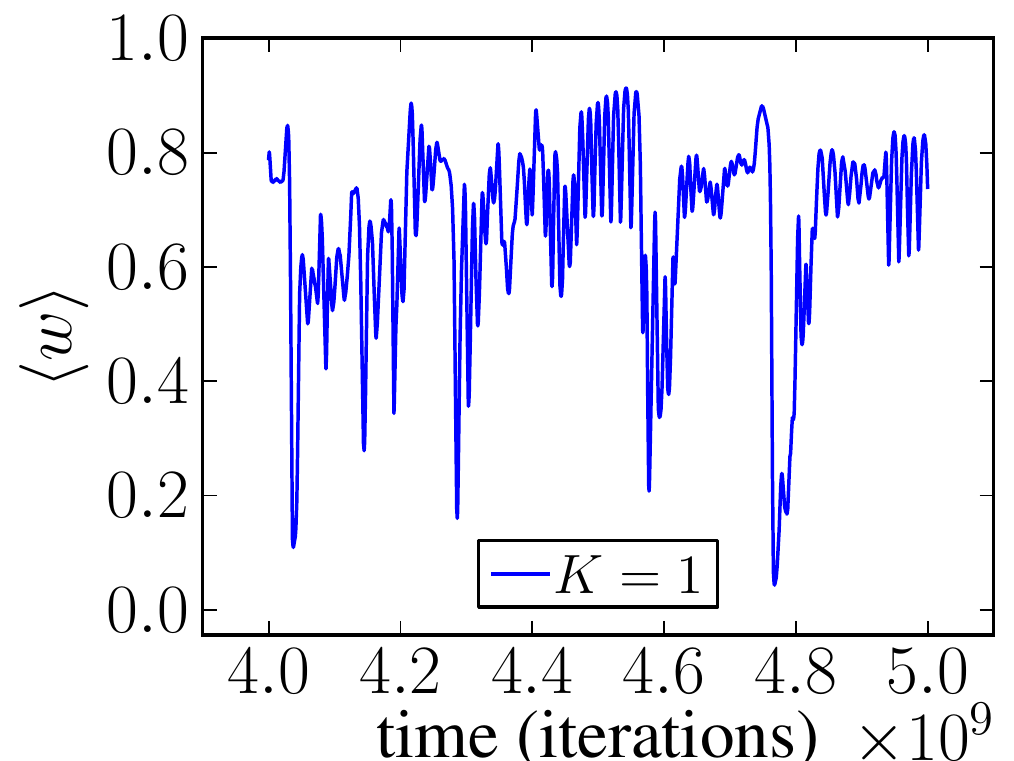}
    (d) $K=1, a > a_c$
  \end{minipage} \caption{Time evolution of the average value for money
  $\avg{w}$, in a population of $N=10^6$ agents with different values of
  $K$ and $a$. The lower right panel (d) shows a zoomed region of the
  lower left panel (c).\label{fig:w-evo}}
\end{figure}

We investigate the dynamics of this model by simulating a population of
$N=10^6$ agents, as well as obtaining some properties analytically in
the limit $N\to\infty$. In order to avoid absorbing states where a
single value for money is fixated on the entire population, we introduce
a small noise probability $r=10^{-6}$ that at each time step a randomly
chosen agent acquires a random value of $w\in[0,1]$. The dynamics will
depend strongly on the parameter $a$, which controls the relative speed
with which rewarders update their strategies, when compared to
donators. If this value is too low, the donators will react fast enough
to changes in available value for money, selecting those with higher $w$
values, and only these agents will have a larger rewarder payoff, and
thus the dynamics will settle on a stable fixed point where the entire
population has the same value of $w$, witch tends asymptotically to one,
as can be seen is Fig.~\ref{fig:w-evo}a. In this situation all rewarders
will on average receive the same number of donators, which will be
distributed according to a Poisson. This is the ideal scenario for
donators, but the worst possible for rewarders, since their average
payoff values reach their maximum and minimum values,
respectively. However, as the remaining panels of Fig.~\ref{fig:w-evo}
show, this situation changes as $a$ is increased. For values of $a>a_c$,
where $a_c$ is a critical value depending on $K$, the average value for
money $\avg{w}$ fluctuates around values smaller than one, since the
rewarders are quick enough to copy low values of $w$, so that there are
few higher values of $w$ left in the population, before the donators
have a chance to react. The values of $w$ remain low since there are no
other options for the donators to choose from. This is an emergent
cartel-like dynamical phase, since all rewarders have settled on a range
of $w$ values which is beneficial to the entire population of rewarders,
and detrimental to the population of donators, which are left with a
restricted choice. However the values of $\avg{w}$ are not quite stable
and fluctuate tremendously, due to influence of the donators and the
always ongoing competition between rewarders.

\begin{figure} 
  \begin{minipage}{0.49\columnwidth}
    \includegraphics[width=\textwidth]{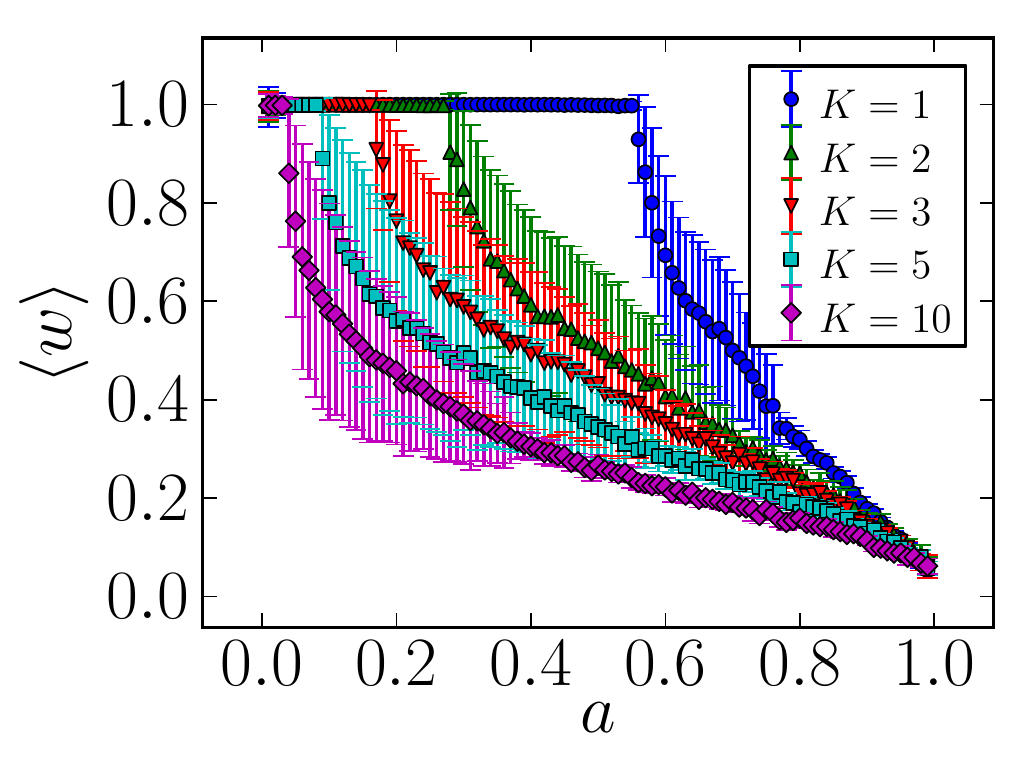}\\
    \flushleft
    \vspace{-11em}
    (a)
    \vspace{8em}
  \end{minipage}
  \begin{minipage}{0.49\columnwidth}
    \includegraphics[width=\textwidth]{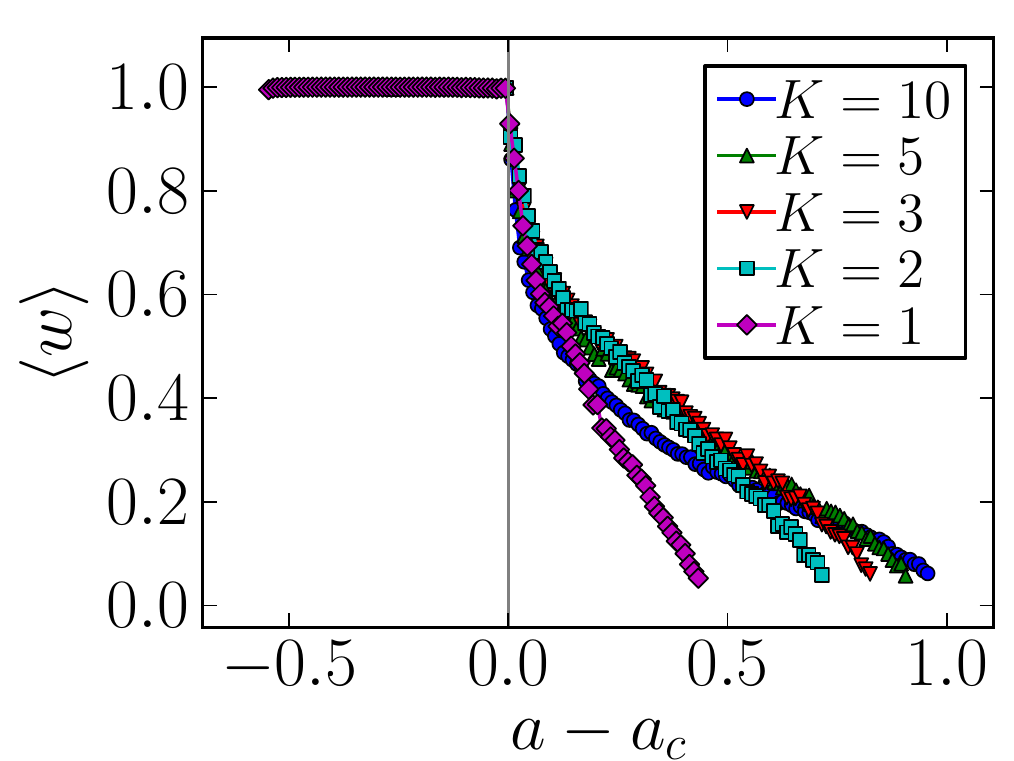}
    \flushleft
    \vspace{-11em}
    (b)
    \vspace{8em}
  \end{minipage} \caption{(a) Phase diagram showing the average value
  for money $\avg{w}$ over time, as a function of $a$, in a population
  of $N=10^6$ agents with different values of $K$. The error bars
  indicate the variance of the time series (not of the average). (b) The
  same curves as in (a) scaled according to the critical value $a_c$,
  calculated from Eq.~\ref{eq:linear}.\label{fig:phase-diag}}
\end{figure}

In Fig.~\ref{fig:phase-diag}a can be seen the phase diagram for diverse
values of $K$, which show the emergence of the cartel phase at a
critical value of $a_c$, after which the $\avg{w} = 1$ ceases to be
stable. The stability of this fixed point can be accessed by a linear
stability analysis: If this fixed point is perturbed by the inclusion of
a small fraction of agents with a lower value of $w<1$, the time
evolution of the probability density $\rho(k, w)$ of agents with this
$w$ value and in-degree $k$ is given by,
\begin{multline}\label{eq:linear}
\frac{\partial}{\partial t}\rho(k, w) \cong a P_0(k) (1-\delta_{k,0})
\sum_{k'>0}\rho(k', w) + \\ (1-a)\left[\frac{k+1}{\avg k} \rho(k+1, w) -
\frac{k}{\avg k} \rho(k, w)\right],
\end{multline}
where $P_0(k)$ is a Poisson distribution with average $K$, and terms of
order $O(\rho(k',w)^2)$ were neglected (see Eq.~\ref{eq:master} below
for the full master equation). The first term of Eq.~\ref{eq:linear}
corresponds simply to the probability of rewarders adopting the invading
strategy, whereas the second term accounts for the probability of agents
with the invading strategy losing donators. Eq.~\ref{eq:linear} is a
linear system, and thus can be written in the form $\dot{\bm{x}} =
M\bm{x}$, where $x_i$ are the individual $\rho(k, w)$ variables, and $M$
is a matrix corresponding to the right-hand side of Eq.~\ref{eq:linear}.
If the value of $a$ is large enough so that the real part of an
eigenvalue of $M$ becomes larger than zero, $\operatorname{Re}\lambda >
0$, the fixed point ceases to be stable. By numerically computing
$\lambda$, one can find the value of $a=a_c$ for which
$\operatorname{Re}\lambda=0$. These values predict exactly the
transition point, as Fig.~\ref{fig:phase-diag}b shows.

\begin{figure} 
  \begin{minipage}{0.49\columnwidth}
    \includegraphics[width=\textwidth]{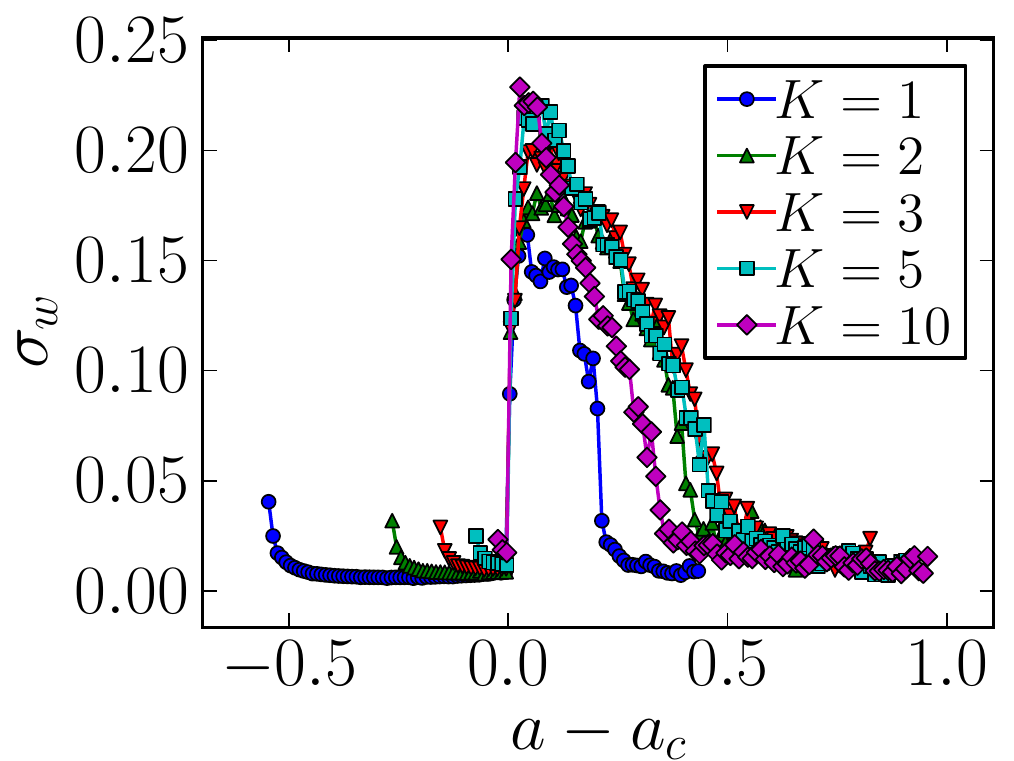}
    \flushleft
    \vspace{-11em}
    \hspace{-0.5em}(a)
    \vspace{9em}
  \end{minipage}
  \begin{minipage}{0.49\columnwidth}
    \includegraphics[width=\textwidth]{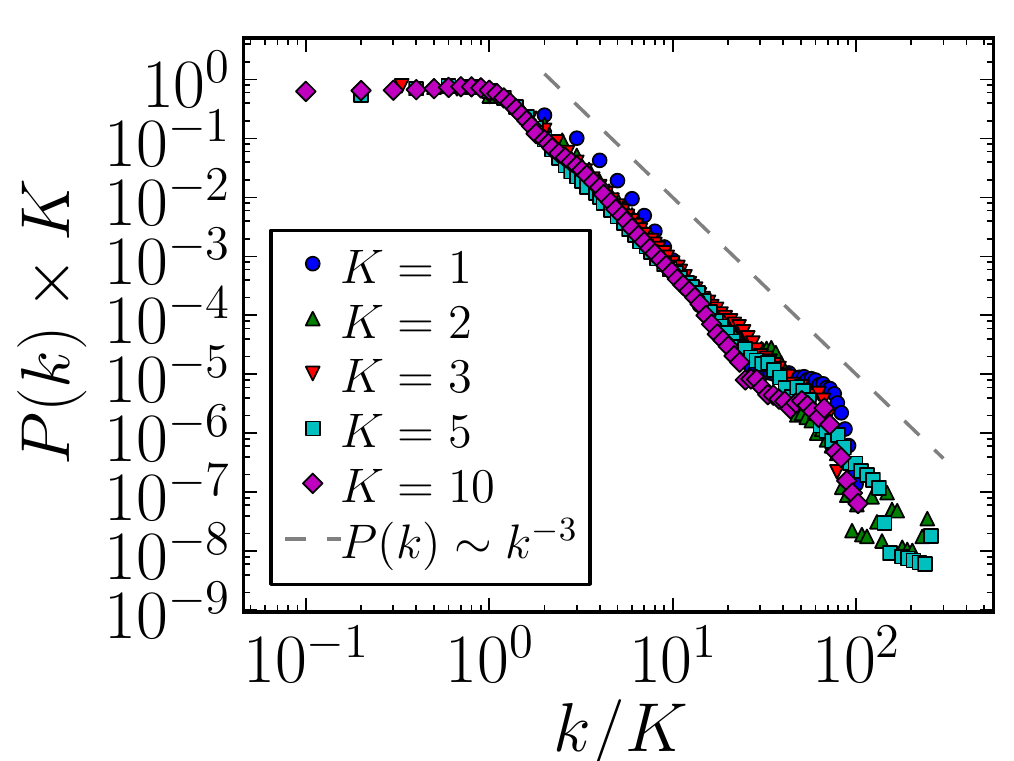}
    \flushleft
    \vspace{-11em}
    (b)
    \vspace{9em}
  \end{minipage}
  \begin{minipage}{0.49\columnwidth}
    \includegraphics[width=\textwidth]{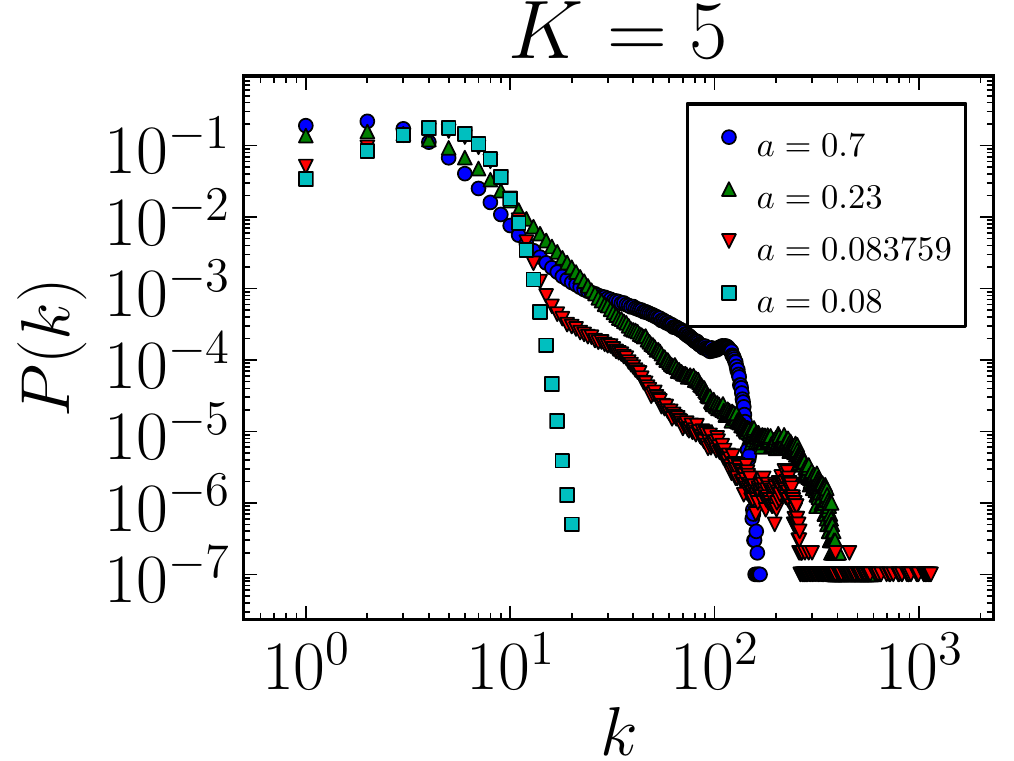}
    \flushleft
    \vspace{-11em}
    (c)
    \vspace{9em}
  \end{minipage}
  \begin{minipage}{0.49\columnwidth}
    \includegraphics[width=\textwidth]{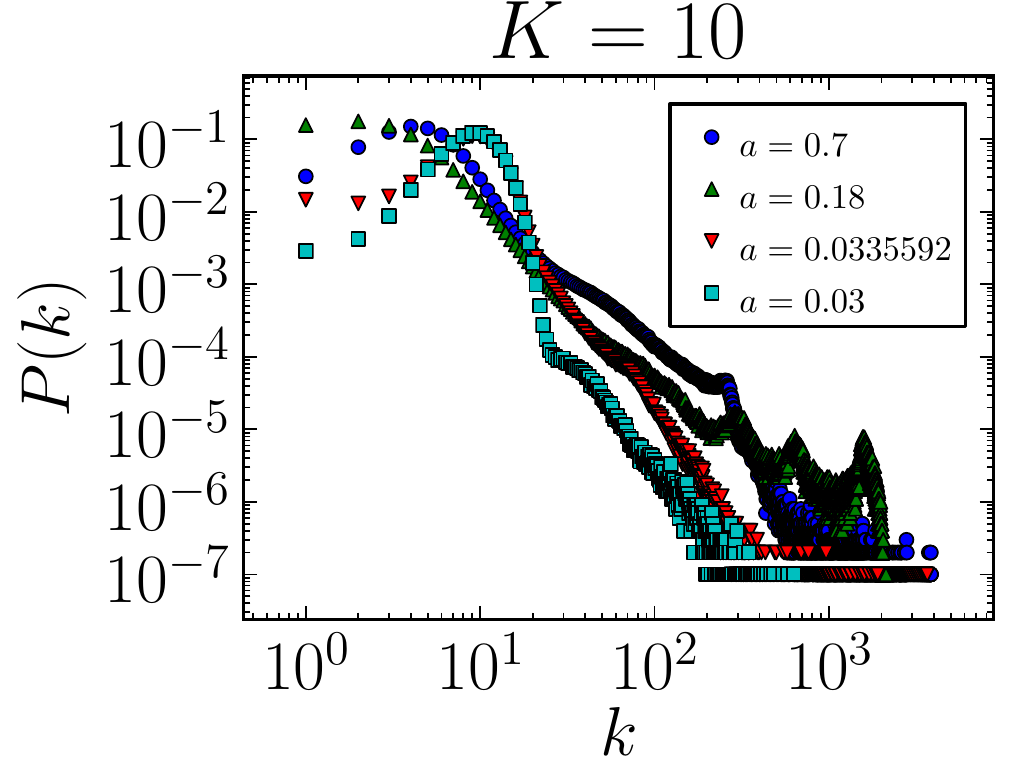}
    \flushleft
    \vspace{-11em}
    (d)
    \vspace{9em}
  \end{minipage}

  \caption{(a) Variance of the average value for money $\avg{w}$ over
  time, as a function of $a$, for different $K$. (b) Rescaled in-degree
  distribution, accumulated over the entire history, for the values of
  $a$ for which $\sigma_\avg{w}$ is maximum. Lower panels: In-degree
  distributions for $K=5$ (c) and $10$ (d), and different values of
  $a$. \label{fig:fluctuations}}
\end{figure}

Exactly at the critical value $a=a_c$, the oscillations of $\avg{w}$
show typical critical behavior, with sharp jumps from the average value
close to one, corresponding to repeated successful invasions of low
value for money, which disappear after a relatively short
time. Interestingly, the variance of the average value over time,
$\sigma_{\avg{w}}$, is largest not exactly at the critical point, but at
values $a>a_c$ close to it, as Fig.~\ref{fig:fluctuations}a
shows. Exactly for the values of $a$ for which $\sigma_{\avg{w}}$ is
maximum, one obtains an universal behavior for the in-degree
distribution of agents (accumulated over the entire history), which
exhibits a power law tail of the form $P(k) \sim k^{-3}$ (see
Fig.~\ref{fig:fluctuations}b). For larger values of $a$ the fluctuations
diminish but remain significant. Indeed the $P(k)$ distributions are
broad for almost the entire parameter space, as the remaining panels of
Fig.~\ref{fig:fluctuations} show.

\begin{figure} 
  \begin{minipage}{0.49\columnwidth}
    \includegraphics[width=\textwidth]{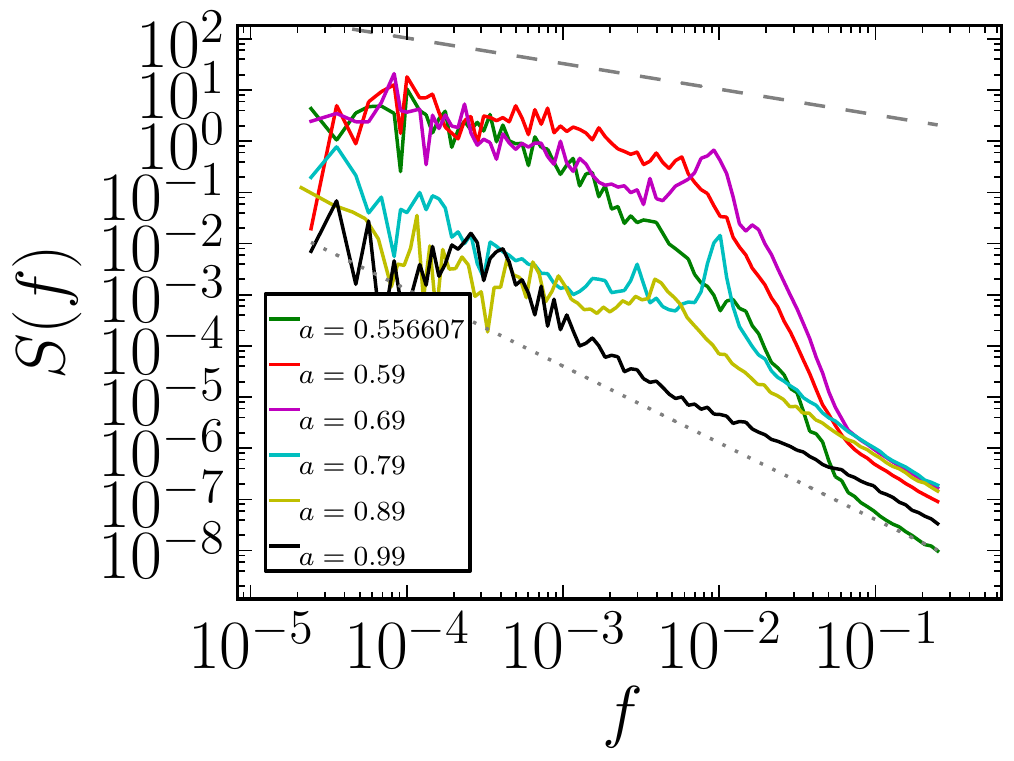}
    \centering
    (a) $K=1$
  \end{minipage}
  \begin{minipage}{0.49\columnwidth}
    \includegraphics[width=\textwidth]{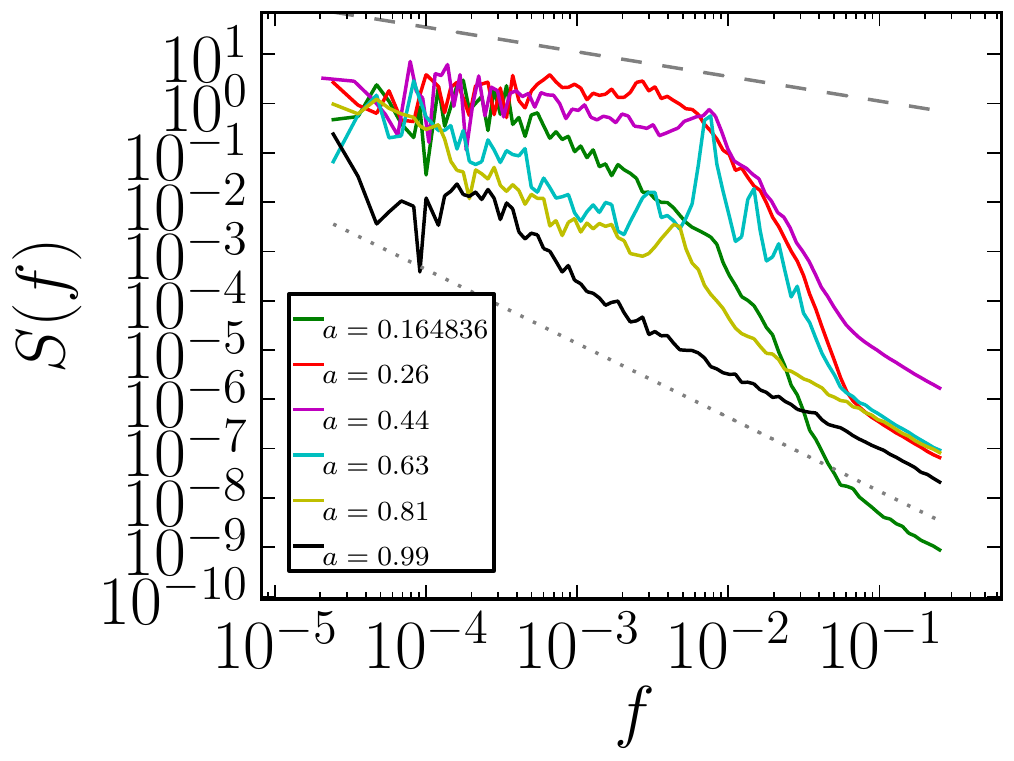}
    \centering
    (b) $K=3$
  \end{minipage}

  \caption{Power spectral density for the time series of $\avg{w}$, and
  different values of $K$ and $a$. The dotted and dashed lines show
  $1/f^\alpha$ curves with $\alpha=3/2$ and $\alpha=1/2$,
  respectively. \label{fig:spectra}}
\end{figure}

A visual analysis of the temporal evolution of the values of $\avg{w}$
cannot reveal the precise characteristics of the fluctuations,
e.g. whether it is aperiodic or quasi-periodic. In
Fig.~\ref{fig:spectra} is shown the spectral density of the time series,
revealing very broad spectra, compatible with aperiodic behavior. For
values of $a$ close to $1$, a $1/f^\alpha$ spectrum is clearly
identified, with $\alpha\cong 3/2$. For lower values of $a$, the
spectrum is divided roughly into lower and high frequency regions, with
stronger fluctuations at lower frequencies. The lower-frequency spectrum
is significantly broad, and is compatible with a $1/f^\alpha$ decay,
with exponents in the range $\alpha \in [3/2, 1/2]$.

\begin{figure} 

  \begin{minipage}{0.32\columnwidth}\centering\includegraphics[width=\columnwidth]{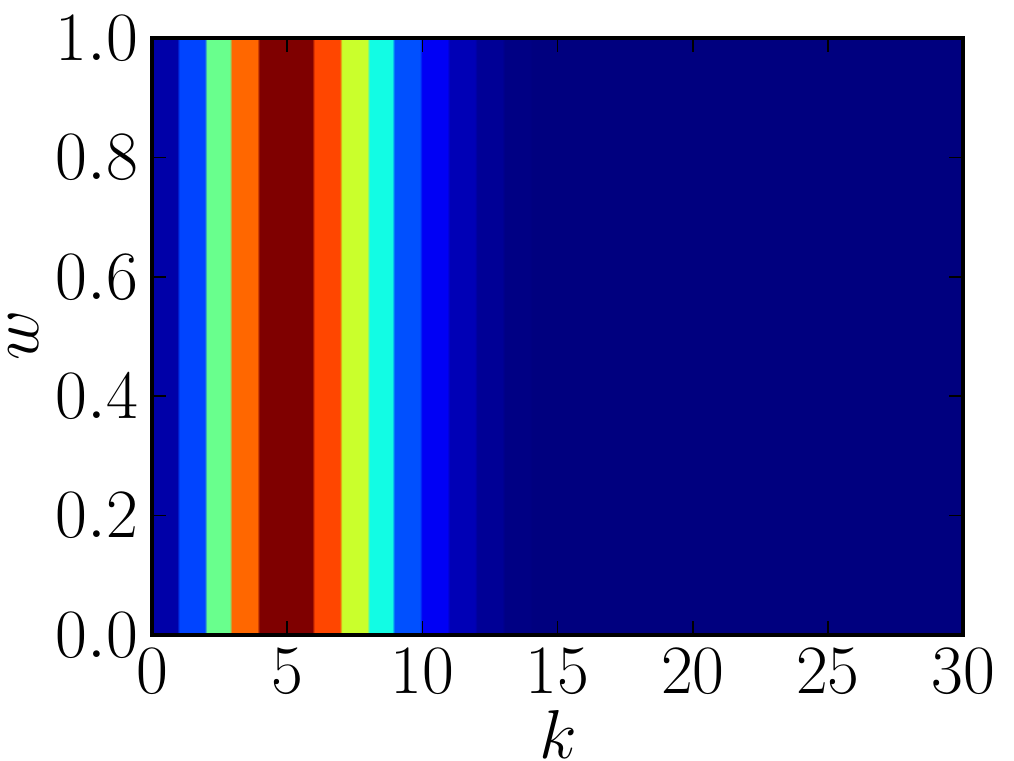}\\
    $t=0$
  \end{minipage}
  \begin{minipage}{0.32\columnwidth}\centering\includegraphics[width=\columnwidth]{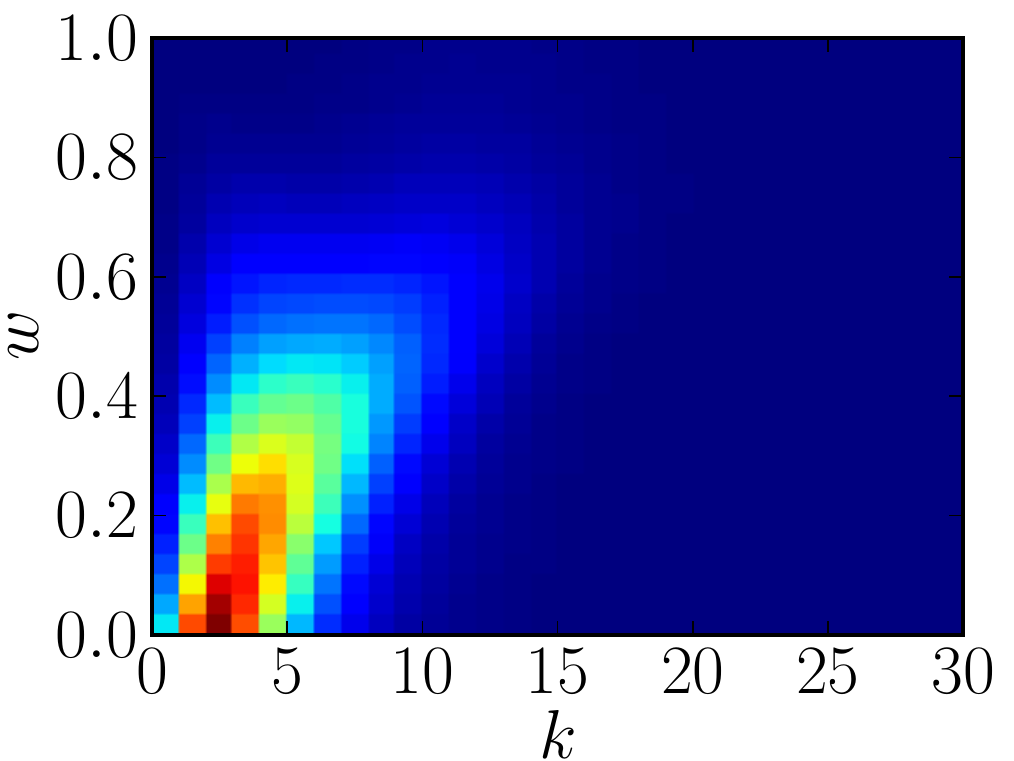}\\
    $t=100$
  \end{minipage}
  \begin{minipage}{0.32\columnwidth}\centering\includegraphics[width=\columnwidth]{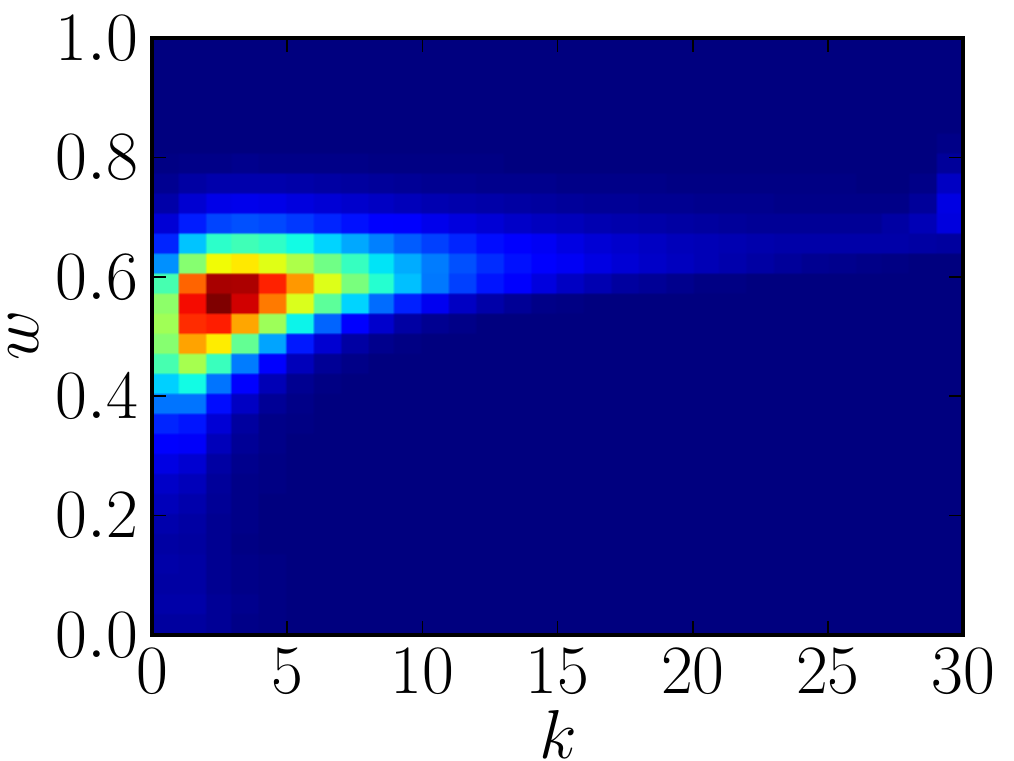}\\
    $t=500$
  \end{minipage}\\
  \begin{minipage}{0.32\columnwidth}\centering\includegraphics[width=\columnwidth]{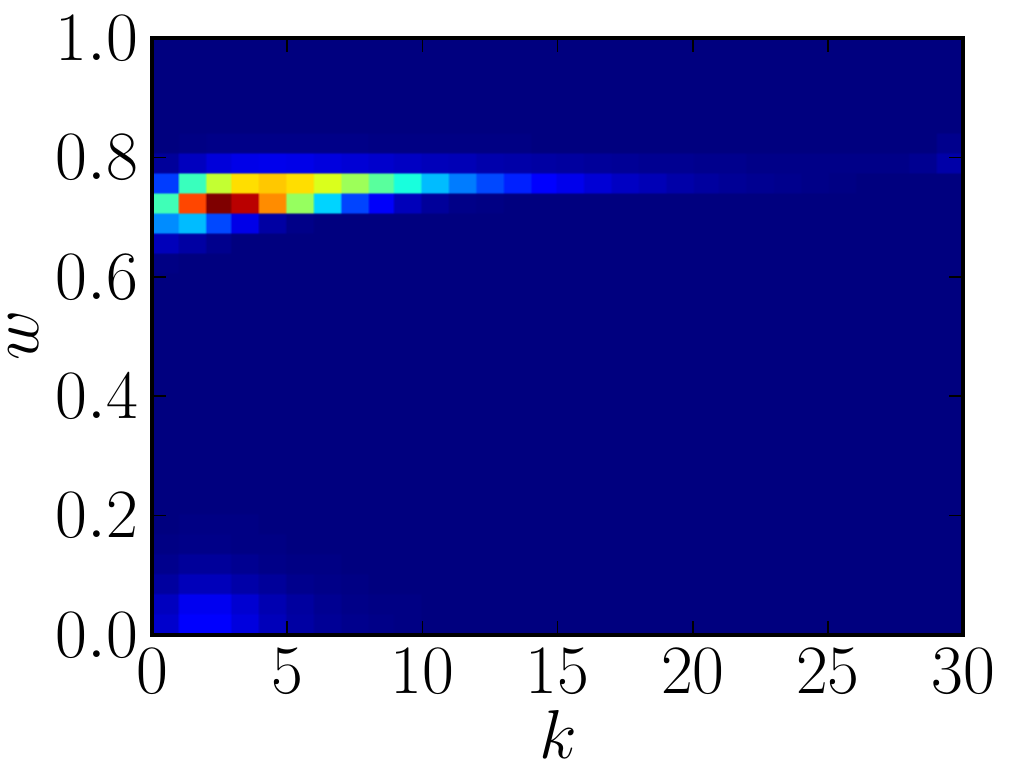}\\
    $t=1000$
  \end{minipage} 
  \begin{minipage}{0.32\columnwidth}\centering\includegraphics[width=\columnwidth]{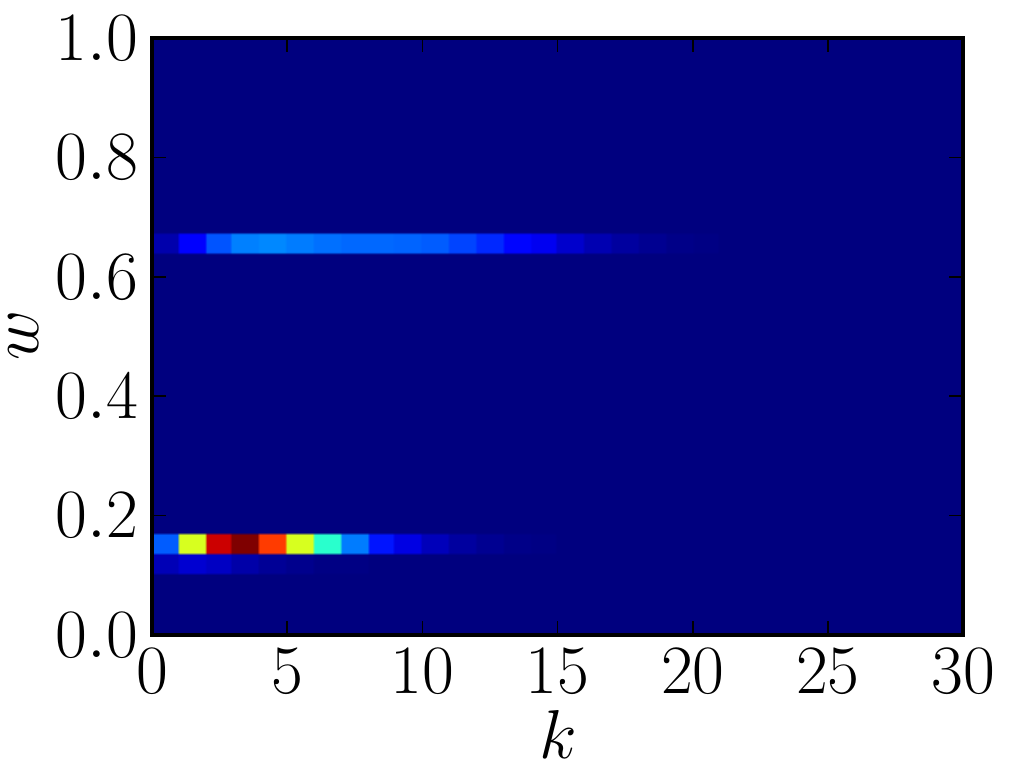}\\
    $t=10000$
  \end{minipage}
  \begin{minipage}{0.32\columnwidth}\centering\includegraphics[width=\columnwidth]{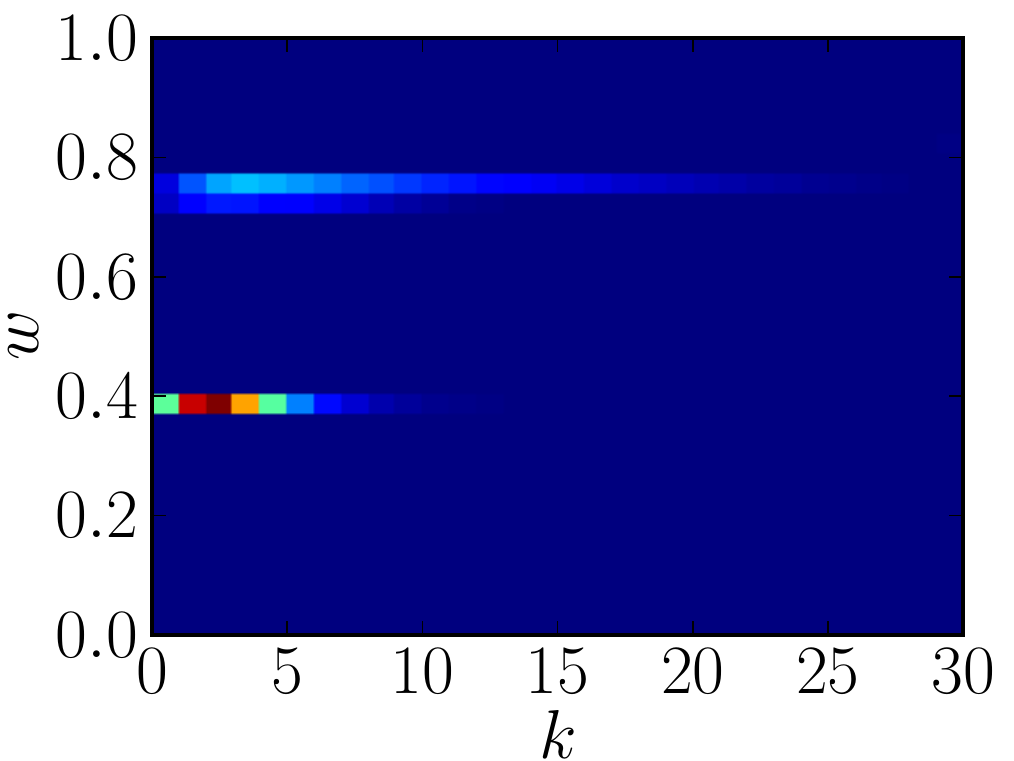}\\
    $t=20000$
  \end{minipage}\\
  \begin{minipage}{0.32\columnwidth}\centering\includegraphics[width=\columnwidth]{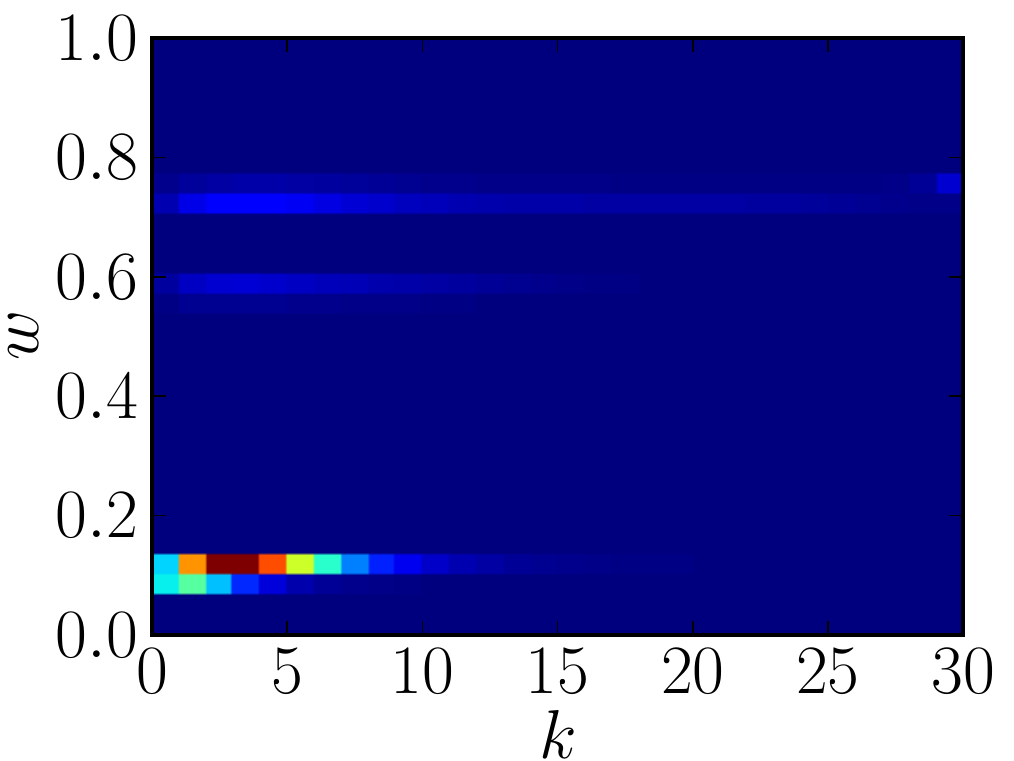}\\
    $t=40000$
  \end{minipage}
  \begin{minipage}{0.32\columnwidth}\centering\includegraphics[width=\columnwidth]{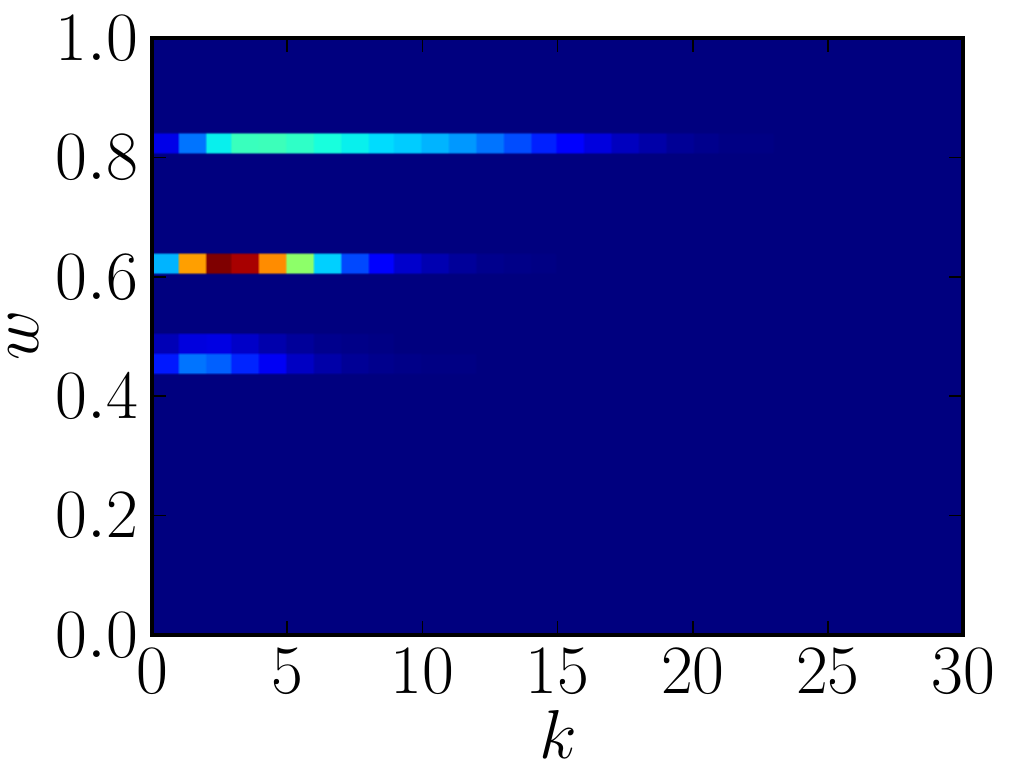}\\
    $t=80000$
  \end{minipage}
  \begin{minipage}{0.32\columnwidth}\centering\includegraphics[width=\columnwidth]{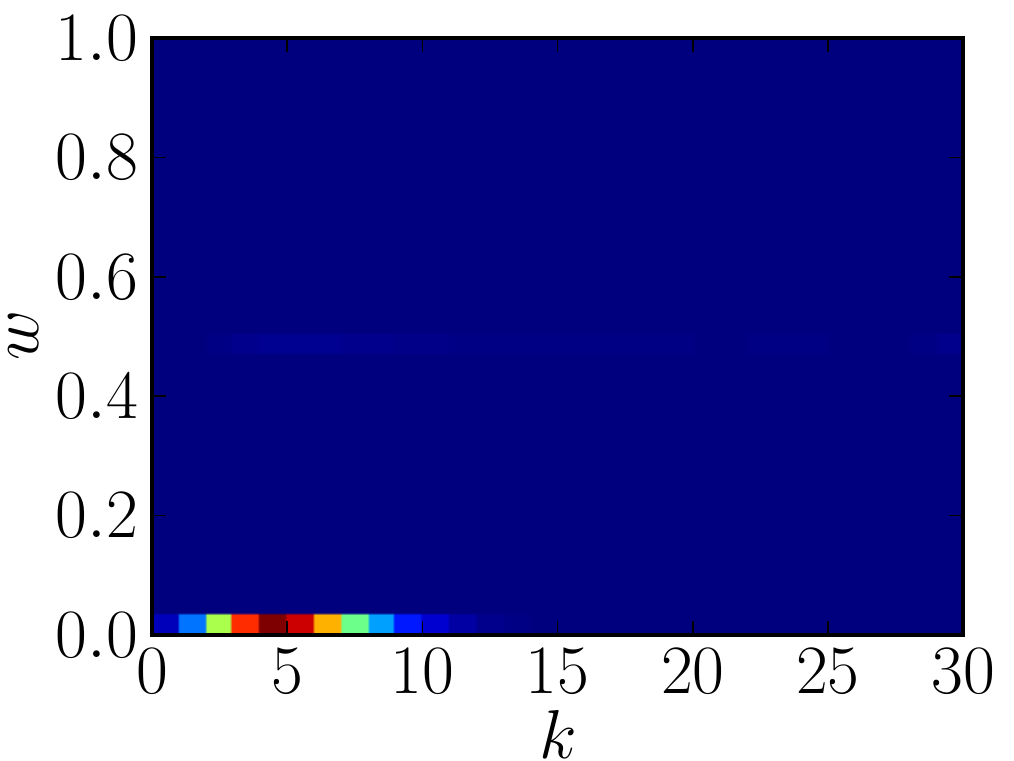}\\
    $t=100000$
  \end{minipage}

  \caption{Temporal evolution of $P(k,w)$, as obtained by integrating
  Eq.~\ref{eq:master}. \label{fig:master}}
\end{figure}

The exact dynamics which give rise to the observed fluctuations can be
explored more closely by specifying the full master equation which
describes the behavior of the system in the limit $N\to\infty$. For
practical reasons, we assume now that the values of $w_i$ must be chosen
from a discrete set of $N_w$ elements distributed uniformly in the $[0,
1]$ range. Taking the limit $N_w\to\infty$, one recovers the exact same
model as before. One can describe the time evolution of the probability
$P(k, w)$ of observing a agent with value for money $w$ and in-degree
$k$ as,
\begin{equation}\label{eq:master}
  \frac{\partial}{\partial t}P(k, w) = a \gamma(k, w) + (1-a) \xi(k,w)
\end{equation}
where $\gamma(k, w)$ and $\xi(k,w)$ describe the rewarder replication
and donator comparison dynamics, respectively. The term $\gamma(k, w)$
is defined as,
\begin{multline}\label{eq:master_reward}
  \gamma(k,w) = \sum_{w'\ne w} P(k, w')\tilde{P}_k\left(k \frac {1-w'}{1-w}, w\right) - \\
  P(k,w)\sum_{w'\ne w}\tilde{P}_k\left(k \frac {1-w}{1-w'}, w'\right),
\end{multline}
where $\tilde{P}_k(k, w) = \sum_{k'\ge k}P(k',w)$. The first term in
Eq.~\ref{eq:master_reward} corresponds to the probability of randomly
selected agent with a lower or equal payoff $k'(1-w')$ selecting $w$ as
the new strategy, and the second term to the probability of an agent
with strategy $w$ finding another agent with higher or equal payoff and
a different value $w'$. The term $\xi(k, w)$ describes the change in
$P(k, w)$ which is due to donators selecting different rewarders, and is
given by,
\begin{multline}\label{eq:master_donate}
  \xi(k,w) = \frac{k + 1}{\avg{k}} P(k+1, w)\left(\tilde{P}_w(w) - P(k,w)\right)\\
  - \frac{k}{\avg{k}} P(k, w)\left(\tilde{P}_w(w) - P(k - 1,w)\right)\\
  + P(k - 1, w)\left(\tilde{P}^*_w(w) - \frac{k}{\avg{k}}P(k,w)\right)\\
  - P(k, w)\left(\tilde{P}^*_w(w) - \frac{k+1}{\avg{k}}P(k+1,w)\right),
\end{multline}
where $\tilde{P}_w(w)=\sum_{k',w'\ge w}P(k', w')$ and
$\tilde{P}^*_w(w)=\sum_{k',w'\ge w}P(k', w')k'/\avg{k}$. The fist two
terms in Eq.~\ref{eq:master_donate} correspond to the probability of a
rewarder with strategy $w$ losing a donator due to a comparison with a
random rewarder with a higher or equal value $w'$. The two remaining
terms describe the converse probability of a rewarder receiving a
donator from another rewarder with value $w'\leq w$.  The time evolution
of $P(k, w)$ for $a>a_c$ is shown in Fig.~\ref{fig:master}, starting
from a random configuration where all values of $w$ are equally
probable, and the in-degree distribution is a Poisson for all values of
$w$. Initially, the mass of the distribution shifts to lower values of
$w$ as agents adopt a value for money with a larger
payoff. Simultaneously, the upper left portion of the distribution
increases in mass, since the rewarders with larger $w$ receive more
donators. Eventually the payoff of the rewarders with high $w$ and $k$
will be large enough to drive the entire distribution upwards. At this
point, all rewarders will receive approximately the same number of
donators, and the system will become susceptible to an invasion of low
value for money. Due to the same dynamics as before, the new front of
low value for money will move upwards in the $w$ axis, prompting the
eventual appearance of yet another front, and so on. Although this
corresponds to cycles of average value for money, the whole dynamics is
aperiodic, and is not easy to predict when the next front will come, and
how it will interact with the preceding ones. This dynamics of
succeeding fronts proceeds indefinitely, and the system never settles on
a fixed point.

In conclusion, we have developed a minimal model of a trust game played
on a population of agents, which displays an emergent cartel-like
behavior, with large fluctuations of value for money. As mentioned
previously, this model provides a qualitative explanation for the price
fluctuation of certain commodity prices, such as gasoline. As many
empirical studies have shown~\cite{eckert_empirical_2011}, the price of
gasoline fluctuates between gas stations. The average price for a given
city often exhibits daily variations, and sometimes fluctuates within
the same day. The average price often rises very fast, and decay more
slowly. This type of oscillation is called Edgeworth price
cycles~\cite{maskin_theory_1988}, and is predicted by simple models
involving two companies, which change their strategy at each round,
reacting to the strategy played by the other company at the previous
round~\cite{maskin_theory_1988, noel_edgeworth_2007}. This model,
however, is not applicable in situations involving many companies. Our
model not only assumes that there are many sellers, but also it
incorporates the behavior of the buyers explicitly. The resulting
oscillations which we observe are a result of the competition in the
entire market, not the steady state behavior of very few companies which
observe each other directly. Furthermore, it sheds light on the question
of market regulation. It is often discussed if the observed fluctuations
are a result of collusion among the gas companies, who attempt to
increase the gas prices in unison~\cite{eckert_empirical_2011}. Although
this is certainly a possibility, our model shows that an explicit
coordinated behavior among sellers is not an indispensable requirement
for a cartel-like behavior.

We acknowledge the support of the DFG under contract INST 144/242-1
FUGG.

\bibliographystyle{apsrev4-1}
\bibliography{bib}
\end{document}